\documentclass[
aps,prd,
12pt,
nopreprintnumbers,
showpacs,
eqsecnum,
nofootinbib
]{revtex4-1}

\usepackage{graphicx}
\usepackage{amssymb}

\begin{document}

\title{Generic Scalar Potentials in Geometric Scalar Gravity}
\author{Nahomi Kan}\email[]{kan@gifu-nct.ac.jp}
\affiliation{National Institute of Technology, Gifu College,
Motosu-shi, Gifu 501-0495, Japan
}
\author{Kiyoshi Shiraishi}\email[]{shiraish@yamaguchi-u.ac.jp}
\affiliation{
Graduate School of Science and Engineering, Yamaguchi University,
Yamaguchi-shi, Yamaguchi 753--8512, Japan}
\date{\today}

\begin{abstract}
We discuss a generic form of the scalar potential appearing in
the geometric scalar theory of gravity. We find the conditions on the
potential by considering weak and strong gravity. The modified
black hole solutions are obtained for generic potentials and the inverse
problems on a black hole and on a spherical body (`pseudo-gravastar') are
investigated.
\end{abstract}



\maketitle

\section{Introduction}
\label{sec1}

General Relativity (GR) is a unique theory of gravity which passes
various experimental tests, at least to our present knowledge.
The modifications of GR are, however, studied by many researchers
for the reason that they may resolve the cosmological riddles about
dark matter and/or dark energy \cite{MG1,MG2,MG3}.

In recent years, Novello and collaborators advocated a novel theory
of gravity, called Geometric Scalar Gravity (GSG) 
\cite{NBMGST,NB,BMNT,BNMGST,Thesis}. In this
theory, the dynamic field is a single scalar field, but a normalized
derivative of the scalar field expresses a part
of the dynamical metric, as well as the scalar field itself. Therefore,
it is possible to avoid the difficulty in old scalar
gravitation theories not including derivative terms
\cite{Kraichman,FN,DH,ST}\footnote{For older many works, please see the
references therein.}.

Novello is also insisting \cite{Novello} that more scalar degrees of
freedom are necessary to explain the gravitational field around the
spinning source more practically. However, it is a challenging problem for
theorists to clarify how GSG, as the simplest model or as a case where
many scalar degrees of freedom freeze, can be adjusted to
realistic gravity.

As a modified theory of gravity, mimetic gravity \cite{mimetic} has been
proposed, where a normalized scalar derivative term is also used for the
dynamical metric.  If one would construct such a modified
theory of gravity from tensor and scalar fields, it is worth examining the
behavior of the scalar model in the limit of a few degree of freedom
and pursuing how many aspects can be found through the study of the
simplest theory. Based on the research, we could advance the study of
theories including a vector field, such as TeVeS \cite{TeVeS}, and more
complicated theories.

Novello and collaborators chose the specific form for the potential of the
scalar field from which the Schwarzschild spacetime is derived as a
spherically symmetric vacuum solution, and further advanced towards the
discussion of the (exotic) cosmology based on it
\cite{NBMGST,NB,BMNT,BNMGST,Thesis}. However, GSG is not considered as a
complete theory, so  we need not require perfect coincidence of the
solution with an exact solution of GR. Actually, we have to
check observational validity through description of the parametrized
post-Newton treatment of the weak gravitational field (see for example,
\cite{Weinberg,Hartle}).  Therefore, there may be a finite ambiguity in
the form of the permissible potential. At first we pay attention to this
point and expose the necessary condition that the potential should
satisfy.

Another important issue of the non-linear theory of gravity is
avoidance of the singularity.  A singular point is concealed by
a horizon of  a black hole in GR. There is room of argument
for whether GSG can also be used without a
correction in the  strong gravity,  but we wish to know what
kind of the potential provides a spherically symmetric black hole solution
with a horizon in GSG.

From these two discussions about weak and strong gravity, we consider
possible forms of the scalar potential and some restrictions on GSG.

This paper is organized as follows.
In Sec.~\ref{sec2}, We begin by reviewing GSG to make this paper
self-contained. Then, we consider a spherical gravitational field in GSG
and clarify the condition on the relation between the scalar potential and
parametrized post-Newtonian description in Sec.~\ref{sec3}.
A possible singularity in the spherical vacuum solution in GSG is studied
in Sec.~\ref{sec4}. The condition on the scalar potential for emerging a
horizon is revealed. In Sec.~\ref{sec5}, the potential which leads to a
slightly modified Schwarzschild solution is obtained, such as an inverse
problem.  In this section, we also consider the
spherical body and investigate behaviors of the energy density and the
pressure of the static fluid for a given de Sitter-like spacetime as
another inverse problem. Finally, we give summary and prospects in
Sec.~\ref{sec7}.

\section{A brief review of GSG}
\label{sec2}
In this section, we briefly review  GSG
\cite{NBMGST}  to make the present paper self-contained.
In GSG, the physical metric $q_{\mu\nu}$ is described by a scalar field
$\Phi$ as
\begin{equation}
q_{\mu\nu}=e^{2\Phi}\left[\eta_{\mu\nu}-
\frac{e^{-4\Phi}V(\Phi)-1}{e^{-4\Phi}V(\Phi)}\frac{\partial_\mu\Phi\partial_\nu\Phi}{w}
\right]\,,
\end{equation}
where
\begin{equation}
w\equiv\eta^{\mu\nu}\partial_\mu\Phi\partial_\nu\Phi\,,
\end{equation}
and $\eta_{\mu\nu}$ is a flat Minkowski metric with the signature
$(+---)$.
The inverse of the metric is then written as
\begin{equation}
q^{\mu\nu}=e^{-2\Phi}\left[\eta^{\mu\nu}+\frac{e^{-4\Phi}V(\Phi)-1}{w}
\eta^{\mu\rho}\eta^{\nu\sigma}\partial_\rho\Phi\partial_\sigma\Phi\right]
\,,
\end{equation}
and note that
\begin{equation}
\sqrt{-q}=\sqrt{-\det
q_{\mu\nu}}=\frac{e^{6\Phi}}{\sqrt{V(\Phi)}}\sqrt{-\eta}\,.
\end{equation}
We should also note that
$q^{\mu\nu}\partial_\nu\Phi=e^{-6\Phi}V(\Phi)\eta^{\mu\nu}\partial_\nu\Phi$.

We call $V(\Phi)$ the (scalar) potential in the present paper.
This function determines the dynamics of the scalar $\Phi$, and only the
property of gravity constrains its concrete form,
as far as we do not assume any symmetry argument.

The action governing the dynamics of the scalar field $\Phi$ with a
potential $V(\Phi)$ is given by
\begin{equation}
S_\Phi=\frac{1}{\kappa}\int\sqrt{-q}\sqrt{V(\Phi)}
q^{\mu\nu}\partial_\mu\Phi\partial_\nu\Phi
d^4x\,,
\end{equation}
where $\kappa$ is a constant.
The variation of the action with respect to the scalar $\Phi$ is
calculated as
\begin{equation}
\delta
S_\Phi=-\frac{2}{\kappa}\int\sqrt{-q}\sqrt{V(\Phi)}(\Box\Phi)\delta\Phi
d^4x\,,
\end{equation}
where
\begin{equation}
\Box\Phi\equiv
\frac{1}{\sqrt{-q}}\partial_\mu(\sqrt{-q}q^{\mu\nu}\partial_\nu\Phi)=
e^{-6\Phi}V(\Phi)\left[\frac{1}{\sqrt{-\eta}}\partial_\mu(\sqrt{-\eta}\eta^{\mu\nu}
\partial_\nu\Phi)+\frac{w}{2}\frac{d}{d\Phi}\ln V(\Phi)\right]\,.
\end{equation}

The action for matter fields is presumed as
\begin{equation}
S_m=\int\sqrt{-q}{\cal L}_m d^4x\,,
\end{equation}
and its variation with respect to the metric is written as follows:
\begin{equation}
\delta S_m=-\frac{1}{2}\int\sqrt{-q}\,T^{\mu\nu}\delta q_{\mu\nu} d^4x\,,
\end{equation}
where
\begin{equation}
T_{\mu\nu}\equiv\frac{2}{\sqrt{-q}}\frac{\partial\sqrt{-q}{\cal
L}_m}{\partial q^{\mu\nu}}
\end{equation}
is the energy-momentum tensor of matter fields.
Some manipulation leads to the following result on the variation with
respect to the scalar $\Phi$:
\begin{equation}
T^{\mu\nu}\delta q_{\mu\nu}=\left[2T+
\left(4-\frac{1}{V}\frac{dV}{d\Phi}\right)E\right]\delta\Phi
-2C^\lambda\partial_\lambda\delta\Phi\,,
\end{equation}
where
\begin{equation}
T=T^{\mu\nu}q_{\mu\nu}\,,\quad 
E=\frac{T^{\mu\nu}\partial_\mu\Phi\partial_\nu\Phi}{\Omega}\,,\quad
C^\lambda=\frac{e^{-4\Phi}V-1}{\Omega}(T^{\lambda\nu}-E q^{\lambda\nu})
\partial_\nu\Phi\,,
\label{ETC}
\end{equation}
with
\begin{equation}
\Omega\equiv q^{\mu\nu}\partial_\mu\Phi\partial_\nu\Phi=e^{-6\Phi}Vw\,.
\end{equation}
We then obtain the variation of the action for matter fields with respect
to $\Phi$ as
\begin{equation}
\delta
S_m=-\int\left[T+\left(2-\frac{1}{2V}\frac{dV}{d\Phi}\right)E+
\nabla_\lambda C^\lambda
\right]\delta\Phi\sqrt{-q}d^4x\,,
\end{equation}
where
\begin{equation}
\nabla_\lambda
C^\lambda\equiv\frac{1}{\sqrt{-q}}(\partial_\lambda\sqrt{-q}\,C^\lambda)\,.
\end{equation}

Now we define the total action of the gravitating system as
\begin{equation}
S=S_\phi+S_m\,.
\end{equation}
The equation derived from this total action can be written as
\begin{equation}
\sqrt{V(\Phi)}\Box\Phi=\kappa\chi\,,
\label{DE}
\end{equation}
where
\begin{equation}
\chi=-\frac{1}{2}\left[T+\left(2-\frac{1}{2V}\frac{dV}{d\Phi}\right)E+
\nabla_\lambda C^\lambda
\right]\,.
\label{source}
\end{equation}

Let us consider the Newtonian limit. The gravitating source can be
estimated as
\begin{equation}
T_{00}\approx \rho\,,
\end{equation}
where $\rho$ is the
energy density and the other components of the energy-momentum
tensor $T_{\mu\nu}$ can be neglected.  

Novello et al.~then assumed
\begin{equation}
q_{00}=e^{2\Phi}\approx 1+2\Phi_N\,,
\end{equation}
where $\Phi_N$ denotes the static Newtonian potential. The other elements
of the metric tensor are taken as those of the Minkowski metric. 

We would like to consider the case with the non-zero asymptotic value of
the scalar field
$\Phi_\infty$. The redefinition of the time coordinate yields the metric
component $q_{00}=e^{2\Phi_\infty}\rightarrow 1$ in the asymptotic region.
In this case, since $|\Phi-\Phi_\infty|\approx
|\Phi_N|\ll 1$, we find that the dynamical equation (\ref{DE}) can be
approximated as
\begin{equation}
\sqrt{V(\Phi_\infty)}\nabla^2\Phi=\frac{\kappa}{2}\rho\,.
\end{equation}
Thus, if we set, with the definition of the Newton constant $G$
\begin{equation}
\frac{\kappa}{\sqrt{V(\Phi_\infty)}}=8\pi G\,,
\end{equation}
we have a correct Newtonian limit.

\section{A static spherical solution and the post-Newtonian parameters}
\label{sec3}
Novello et al.~considered a static spherical solution in GSG
\cite{NBMGST}. We are interested in generic spherically symmetric
solutions of this theory. We start with their argument in this section.
First, they considered the flat metric in the spherical coordinates,
\begin{equation}
\eta_{\mu\nu}dx^\mu dx^\nu=dT^2-dR^2-R^2d\Omega^2\,,
\end{equation}
with $d\Omega^2$ is the line element on a unit sphere.
Because the spherical symmetry enforces that $\Phi=\Phi(R)$,
the physical line element is given as
\begin{equation}
ds^2=q_{\mu\nu}dx^\mu
dx^\nu=e^{2\Phi}dT^2-\frac{e^{6\Phi}}{V(\Phi)}dR^2-e^{2\Phi}{R^2}d\Omega^2\,.
\end{equation}
Now, defining $t\equiv e^{\Phi_\infty}T$ and converting the radial
coordinate to
$r\equiv e^\Phi R$, we find
\begin{equation}
ds^2=B(r)dt^2-A(r)dr^2-r^2d\Omega^2\,,
\label{stm}
\end{equation}
where
\begin{eqnarray}
B(r)&=&e^{2(\Phi-\Phi_\infty)}\,,\\
A(r)&=&\frac{e^{4\Phi}}{V(\Phi)}\left(1-r\frac{d\Phi}{dr}\right)^2\,.
\label{defA}
\end{eqnarray}

To obtain an asymptotically flat spacetime, we assume
\begin{equation}
\lim_{r\rightarrow\infty}\Phi(r)=\Phi_\infty\,,\quad
\lim_{r\rightarrow\infty}r\Phi'(r)=0\,,\quad
e^{-4\Phi_\infty}V(\Phi_\infty)=1\,,
\end{equation}
where $\Phi'(r)$ is the first derivative of $\Phi(r)$.

The equation of motion in vacuum $\Box\Phi=0$ implies
\begin{equation}
\frac{d}{dr}\left[\sqrt{\frac{B(r)}{A(r)}}r^2\frac{d\Phi}{dr}\right]=0
\,.
\label{dyq}
\end{equation}

Novello et al.~\cite{NBMGST} insisted that the form
for $V(\Phi)$ is chosen so as to realize
the exact Schwarzschild solution. So, they chose
\begin{equation}
\Phi_\infty=0\,,\quad
V(\Phi)=V_{\rm
N}(\Phi)=\frac{1}{4}{e^{2\Phi}(1-3e^{2\Phi})^2}\,.
\label{NP}
\end{equation}
This criterion for the choice of $V(\Phi)$ is too severe
for a  model of weak gravity, we think.
The Newtonian limit has been examined in the previous section.
We study here the post-Newtonian limit.
The possible asymptotic behavior determines the form of the scalar field
as
\begin{equation}
\Phi_N\equiv\Phi-\Phi_\infty=-\frac{GM}{r}-g\frac{G^2M^2}{r^2}+O((GM/r)^{3})\,,
\end{equation}
where $M$ is the gravitational mass and $g$ is a constant of 
dimensionless. The potential $V(\Phi)$ is assumed up to the linear order
in
$\Phi_N$, corresponding to the perturbative regime and we set
\begin{equation}
e^{-4\Phi}V(\Phi)=1+4k\Phi_N+O(\Phi_N^2)\,.
\end{equation}

Then the metric components $B(r)$ and $A(r)$ are expressed in terms of
the parameters as
\begin{eqnarray}
B(r)&=&1-\frac{2GM}{r}+\frac{2(1-g)G^2M^2}{r^2}+O((GM/r)^3)\,,\\
A(r)&=&1+\frac{2(2k-1)GM}{r}+O((GM/r)^2)\,.
\end{eqnarray}
and we then find that
\begin{equation}
\sqrt{\frac{B(r)}{A(r)}}r^2\frac{d\Phi}{dr}=GM+\frac{2(g-k)G^2M^2}{r}
+O(GM(GM/r)^2)
\,.
\end{equation}
Therefore up to the order in the present consideration, the dynamic
equation (\ref{dyq}) tells us $g=k$.

According to the metric form with two post-Newtonian parameters
$\beta$ and
$\gamma$ \cite{Weinberg,Hartle}
\begin{eqnarray}
B(r)&=&1-\frac{2GM}{r}+\frac{2(\beta-\gamma)G^2M^2}{r^2}+O((GM/r)^3)\,,\\
A(r)&=&1+\frac{2\gamma GM}{r}+O((GM/r)^2)\,,
\end{eqnarray}
we can express two of the post-Newtonian parameters $\beta$ and
$\gamma$ by using the linear coefficient $k$ in $V(\Phi)$ as
\begin{equation}
\beta=k\,,\quad \gamma=2k-1\,.
\end{equation}
For the experimentally viable values $\beta=\gamma=1$
\cite{Weinberg,Hartle}, we should take
$k=1$. Of course, the expansion of the potential chosen by 
Novello et al.~(\ref{NP}) shows that it is in the case.

To summarize, we should select the form of the scalar potential $V(\Phi)$
as
\begin{equation}
e^{-4\Phi}V(\Phi)=1+4\Phi_N+O(\Phi_N^2)\,,
\label{NPP}
\end{equation}
with $\Phi_N=\Phi-\Phi_\infty$
for viable post-Newtonian parameters $\beta$ and $\gamma$.
The equivalence of the gravitational and the inertial mass of the
spherical gravitating body then holds. 
Note that (\ref{NPP}) is equivalent to
\begin{equation}
\frac{1}{V(\Phi)}\frac{dV(\Phi)}{d\Phi}=8\quad\mbox{at~}\Phi=\Phi_\infty\,.
\label{vv}
\end{equation}



Since a condition for the potential is obtained now,
we also try to investigate the asymptotic gravitational field around an
electric point charge in GSG. 
We introduce a point charge and consider the
spherically-symmetric solution. The Lagrangian density for the
electromagnetic field is given by
\begin{equation}
{\cal L}_{EM}=-\frac{1}{16\pi}F_{\mu\nu}F^{\mu\nu}\,,
\end{equation}
where $F_{\mu\nu}\equiv\partial_\mu A_\nu-\partial_\nu A_\mu$.
The energy-momentum tensor obtained from this Lagrangian is
\begin{equation}
T_{\mu\nu}=-\frac{1}{4\pi}\left(F_{\mu\nu}^2-\frac{1}{4}F^2
q_{\mu\nu}\right)\,,
\end{equation}
where $F^2_{\mu\nu}\equiv F_{\mu\rho}F_{\nu\sigma} q^{\rho\sigma}$
and $F^2\equiv F^2_{\mu\nu} q^{\mu\nu}$.

A spherically-symmetric solution for the equation of motion $\nabla_\mu
F^{\mu\nu}=0$ is given by
\begin{equation}
F_{0r}=\frac{Q}{r^2}\sqrt{B(r)A(r)}\,,
\end{equation}
where $Q$ is a constant corresponding to the electric charge and
we use the standard metric (\ref{stm}).
Then we find
\begin{equation}
T=0\,,\quad E=\frac{1}{8\pi}\frac{Q^2}{r^4}\,,\quad C^\lambda=0\,,
\end{equation}
 in $\chi$ (\ref{source}) in the right-hand side of the
equation for the scalar field (\ref{DE}).
The equation of motion for the scalar field $\Phi(r)$ becomes
\begin{equation}
\sqrt{\frac{V(\Phi(r))}{A(r)B(r)}}\frac{1}{r^2}\frac{d}{dr}
\left[\sqrt{\frac{B(r)}{A(r)}}r^2\frac{d\Phi(r)}{dr}\right]=
\frac{\kappa}{16\pi}\left(2-\frac{1}{2V}\frac{dV}{d\Phi}\right)
\frac{Q^2}{r^4}
\,,
\end{equation}

In the asymptotic region of large $r$, using (\ref{NPP}), we can
approximate the equation as
\begin{equation}
\sqrt{\frac{V(\Phi_\infty)}{A(r)B(r)}}\frac{1}{r^2}\frac{d}{dr}
\left[\sqrt{\frac{B(r)}{A(r)}}r^2\frac{d\Phi(r)}{dr}\right]\approx
-\frac{\kappa}{8\pi}
\frac{Q^2}{r^4}
\,,
\end{equation}
and we can also find an approximate solution
\begin{equation}
B(r)=1-\frac{2GM}{r}-\frac{GQ^2}{r^2}+O((GM/r)^3)\,.
\end{equation}
Unfortunately, this does not coincide with the Reissner-Nordstr\"om
solution in GR, because of the wrong sign in front of the $r^{-2}$ term.
Although we do not pursue the charged solution in GSG further in this
paper, we suppose that the presence of non-minimal couplings
between the vector and scalar fields could change the asymptotic
behavior and it will be worth studying this possibility.

\section{A static spherical vacuum solution and singularity}
\label{sec4}
As we have seen in the previous section, the vacuum solution with
spherical symmetry satisfies
\begin{equation}
\sqrt{\frac{b(r)}{A(r)}}r^2\frac{d\Phi}{dr}=\frac{1}{2}
\frac{1}{\sqrt{A(r)b(r)}}r^2\frac{db(r)}{dr}
=GMe^{\Phi_\infty}>0
\,,
\label{te}
\end{equation}
where $b(r)\equiv e^{2\Phi(r)}$ and $A(r)$ has been defined as
(\ref{defA}). In this section, we start with undetermined asymptotic value
for the scalar, $\Phi_\infty$.
The equation (\ref{te}) means that
$q_{00}=B(r)=b(r) e^{-2\Phi_\infty}$ is a monotonically increasing
function of
$r$. Moreover, if the value of $b$ is positive everywhere, the above
equation can be rewritten as 
\begin{equation}
\frac{\sqrt{b(r) e^{-4\Phi}V(\Phi(r))}}{\left|2-\frac{rb'(r)}{b(r)}
\right|}r^2\frac{b'(r)}{b(r)}=
\frac{\sqrt{V(\Phi(r))/b(r)}}{\left|2b(r)-{rb'(r)}
\right|}r^2{b'(r)}=GMe^{\Phi_\infty}\,,
\label{sg}
\end{equation}
where the prime (${}'$) stands for the derivative with respect
to $r$.

If we assume that $b>0$ and $\sqrt{V(\Phi(r))/b(r)}$ is finite
in the vicinity of $r=0$, we see that
\begin{equation}
\frac{b'(r)}{b(r)}\rightarrow\infty\quad\mbox{as}\quad r\rightarrow 0\,.
\end{equation}
(and also $\Phi'(r)\rightarrow\infty$) from Eq.~(\ref{sg}).
As far as we assume a non-singular potential ($V(\Phi)<\infty$ for
$-\infty<\Phi\le\Phi_\infty$) and non-singular metric $q_{00}\propto b$
for $r>0$, the vacuum solution has a singularity at the origin. This
fact might bring about a possibility that a naked singularity emerges
after a possible gravitational collapse\footnote{The time-dependent
dynamics in GSG has not been studied, however. We should study the
gravitational collapse in GSG in future.}.
The difficulty of obtaining non-singular vacuum solution forces us to
examine the solution of which metric becomes singular at finite $r$,
which can be regarded as a horizon radius.%
\footnote{The existence of naked singularity in nature is
still controversial, however. For example, see Ref.~\cite{CVE,VE,VK}.} 
If we set $b(r_g)=0$ and $b'(r_g)\ne 0$ at the radius (of
infinite-redshift)
$r=r_g>0$, Eq.~(\ref{te}) tells us that
\begin{equation}
0<\lim_{r\rightarrow r_g}\sqrt{A(r)b(r)}<\infty\,.
\end{equation}
This condition is a necessary condition that the radius $r=r_g$ can be the
horizon radius. Then, because $\lim_{r\rightarrow r_g}\Phi(r)=-\infty$,
it should be satisfied that
\begin{equation}
0<\lim_{\Phi\rightarrow -\infty}\sqrt{
e^{-2\Phi}V(\Phi)}<\infty\,.
\end{equation}
 
Further observation leads to that fact that the specific
combination 
$2-\frac{rb'}{b}$ in (\ref{sg}), which comes from $A(r)$, becomes negative
at
$r=r_g$. Since this must be positive for a large $r$, we conclude that
the quantity
$2-\frac{rB'}{B}$ vanishes at a certain radius
$r=r_0>r_g$. This condition requires 
\begin{equation}
V(\Phi(r_0))=0\,,
\label{r0}
\end{equation}
because the both sides of (\ref{sg}) should be finite.
Therefore the differential equation (\ref{sg}) for $b(r)$ is rewritten as
\begin{eqnarray}
\frac{r}{2}\frac{b'(r)}{b(r)}&=&
\left(1-r\frac{\sqrt{e^{-2\Phi(r)}V(\Phi(r))}}{GMe^{\Phi_\infty}}
\right)^{-1}
\quad\mbox{for }r_g<r<r_0\,,\\
\frac{r}{2}\frac{b'(r)}{b(r)}&=&
\left(1+r\frac{\sqrt{e^{-2\Phi(r)}V(\Phi(r))}}{GMe^{\Phi_\infty}}
\right)^{-1}
\quad\mbox{for }r>r_0\,,
\end{eqnarray}

In order to obtain the smooth function (such as $b(r)\in
C^\infty$), it should be assumed that
\begin{equation}
e^{-2\Phi}V(\Phi)=[F(b)]^2\,,
\end{equation}
and $F(b)$ has a vanishing point, according to Eq.~(\ref{r0}). 
Then the horizon radius $r_g$ is determined as
\begin{equation}
\frac{GMe^{\Phi_\infty}}{r_g}=|F(0)|\equiv F_0\,.
\end{equation}

We will examine the case that the equation has exact analytic
solutions. In the simplest case, $F$ is linear in $b$ such that 
\begin{equation}
F(b)=F_0\left(1-f_1 b\right)\,,
\end{equation}
where $F_0$ and $f_1$ are positive constants.
Adopting a new variable $y\equiv 1-r_g/r$, the differential equation
for $b(y)$ becomes
\begin{equation}
\frac{db}{dy}=\frac{2 b}{-y+1-F(b)/F_0}=\frac{2 b}{-y+f_1 b}\,,
\end{equation}
and the boundary condition is $b(0)=0$.
The solution is
\begin{equation}
b=\frac{3}{f_1}y=\frac{3}{f_1}\left(1-\frac{r_g}{r}\right)\,.
\end{equation}
Then, we find $e^{2\Phi_\infty}=b(1)=\frac{3}{f_1}$ and
\begin{eqnarray}
B(r)&=&e^{2(\Phi-\Phi_\infty)}=1-\frac{r_g}{r}\,,\\
A(r)&=&\frac{3}{4F_0^2f_1}\left(1-\frac{r_g}{r}\right)^{-1}\,.
\end{eqnarray}
To obtain asymptotic flat spacetime, we must choose
\begin{equation}
F_0^2=\frac{3}{4f_1}\,.
\end{equation}
Then, the scalar potential has the form
\begin{equation}
V(\Phi)=\frac{3}{4f_1}e^{2\Phi}\left(1-f_1 e^{2\Phi}\right)^2
=\frac{e^{4\Phi_\infty}}{4}e^{2(\Phi-\Phi_\infty)}\left[1-3
e^{2(\Phi-\Phi_\infty)}\right]^2\,,
\label{gpot}
\end{equation}
and $V(\Phi_\infty)=e^{4\Phi_\infty}$. Thus, the post-Newtonian
condition (\ref{NPP}) is satisfied.
The spacetime is the Schwarzschild spacetime, and thus $\beta=\gamma=1$.
The Newton limit also tells $r_g=2GM$, where
$G=\kappa/(8\pi\sqrt{V(\Phi_\infty)})=f_1\kappa/(24\pi)$.
The potential (\ref{gpot}) is essentially the Novello's potential
$V_{\rm N}(\Phi)$, provided that $f_1=3$ (which implies $\Phi_\infty=0$).

Incidentally, choosing the function $F(b)$ as
\begin{equation}
F(b)=F_0\left(1-f_p b^p\right)\,,
\end{equation}
where $f_p$ is a constant, the differential equation for $b(y)$ becomes
\begin{equation}
\frac{db}{dy}=\frac{2 b}{-y+f_p b^p}\,,
\end{equation}
and then the equation is analytically solved as
\begin{equation}
b=\left(\frac{2p+1}{f_p}\right)^{1/p}y^{1/p}=\left(\frac{2p+1}{f_p}\right)^{1/p}\left(1-
\frac{r_g}{r}\right)^{1/p}\,,
\end{equation}
and  $e^{2\Phi_\infty}=\left(\frac{2p+1}{f_p}\right)^{1/p}$.
The components of the metric are then
\begin{eqnarray}
B(r)&=&e^{2(\Phi-\Phi_\infty)}=\left(1-\frac{r_g}{r}\right)^{1/p}\,,\\
A(r)&=&\frac{1}{4pF_0^2}\left(\frac{2p+1}{f_p}\right)^{1/p}\left(1-\frac{r_g}{r}
\right)^{-2+1/p}\,.
\end{eqnarray}
If we choose $\frac{1}{4pF_0^2}\left(\frac{2p+1}{f_p}\right)^{1/p}=1$,
the solution describes an asymptotically flat spacetime, i.e.,
\begin{equation}
ds^2=\left(1-\frac{r_g}{r}\right)^{1/p}dt^2-\left(1-\frac{r_g}{r}
\right)^{-2+1/p}-r^2d\Omega^2\,.
\label{csol}
\end{equation}
The location $r=r_g$ is, however, a singularity of the metric
(\ref{csol}), which comes from the fact that $db/dr|_{r=r_g}$ does not
take a finite value (i.e., $b'(0)=0$ or $b'(0)=\infty$, where $b'(y)$ is
the first derivative of $b(y)$). 

The physical condition $\beta=\gamma$ adopted in the previous section,
i.e., the absence of
$1/r^2$ term in the expansion of
$B(r)$ in terms of powers of $1/r$, is transformed as $\lim_{y\rightarrow
1}b''(y)=0$, where $b''(y)$ is the second derivative of $b(y)$.
Moreover, we can rewrite the condition (\ref{vv}), by using the
differential equation for $b(y)$, as
\begin{equation}
2b(1)F'(b(1))=3F(b(1))\,,
\end{equation}
where $F'(b)$ is the first derivative of $F(b)$.

We further quest another possible potential,
which leads to the existence of a horizon as well as the nice behavior at
weak gravity, $\beta=\gamma=1$. We assume, with constants $f_1$, $f_2$
and $f_3$,
\begin{equation}
F(b)=F_0\left(1-f_1 b-f_2b^2-f_3b^3\right)\,.
\end{equation}

\begin{figure}[ht]
\centering
\includegraphics[height=3.5cm]
{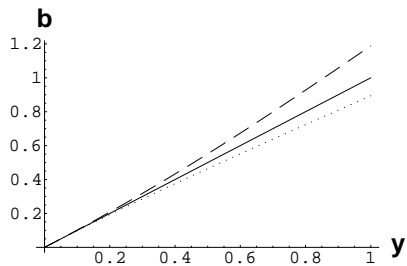}
\caption{%
The solutions for $b(y)$ are plotted against $y$.
The rigid line exhibits the case with $f_1=3$, $f_2=0$ and $f_3=0$.
The dotted line  exhibits the case with $f_1=3$,
$f_2=1$ and $f_3\approx -0.52226$.
The broken line indicates the case with $f_1=3$, $f_2=-1$ and
$f_3\approx 0.39273$.}
\label{fig1}
\end{figure}

To satisfy $\lim_{y\rightarrow 1}b''(y)=0$, we must choose a set of
parameters, $f_1$, $f_2$ and $f_3$.
A set of parameters can be found numerically as
$f_1=3$, $f_2=1$ and $f_3\approx -0.52226$. Then, $b(1)\approx 0.893544$
and
$F_0\approx 0.448751$. Therefore, the horizon radius is $r_g\approx
2.10645 GM$.
Another set is
$f_1=3$, $f_2=-1$ and $f_3\approx 0.39273$. Then, $b(1)\approx 1.18827$.
and $F_0\approx 0.601671$. The horizon radius is $r_g\approx 1.81175
GM$ in this case.
The solutions for $b(y)$ are displayed in FIG.~\ref{fig1}.
We find that the horizon radius can become larger or smaller than $2GM$ by
selecting the parameters, i.e., the form of the scalar potential
$V(\Phi)$.

\section{Inverse problems in GSG}
\label{sec5}

\subsection{An inverse problem on black holes}
\label{5_1}
By now, we have found the conditions for the scalar potential $V(\Phi)$.
Vast ambiguity in the form of the potential remains with the
conditions. In this section, we consider an `inverse problem',
which require the potential form to obtain a given simple metric
component. We assume the simplest `solution' which is permissible up to
the post-Newtonian order,
\begin{equation}
b=\frac{3}{f}\left[y+h\left(y^2-\frac{1}{3}y^3\right)\right]\,,
\end{equation}
where $f$ $(>0)$ and $h$ $(>-1)$ are constants.
Since $y=1-{r_g}/{r}$, we have
\begin{equation}
b=e^{2\Phi}=e^{2\Phi_\infty}\left[1-\frac{2GM}{r}+
\frac{4}{3}\delta\frac{(2GM)^3}{r^3}\right]\,,
\label{eqr}
\end{equation}
where
\begin{equation}
e^{2\Phi_\infty}=\frac{3}{f}\left(1+\frac{2}{3}h\right)\,,\quad
\delta=\frac{h}{4}\frac{\left(1+\frac{2}{3}h\right)^2}{(1+h)^3}\,,
\end{equation}
\begin{equation}
\frac{r_g}{2GM}=\frac{e^{\Phi_\infty}}{2F_0}=\frac{1+
\frac{2}{3}h}{1+h}\,,\quad
F_0^2=\frac{3}{4f}\frac{(1+h)^2}{1+\frac{2}{3}h}\,.
\end{equation}
If we solve the equation (\ref{eqr}) for $2GM/r$, we find
\begin{equation}
\frac{2GM}{r}=\frac{1}{\sqrt{\delta}}\sin\left[
\frac{1}{3}\arcsin\left\{3\sqrt{\delta}(1-e^{-2\Phi_\infty}b)\right\}
\right]\quad\mbox{for~}\delta>0\,,
\end{equation}
\begin{equation}
\frac{2GM}{r}=\frac{1}{\sqrt{|\delta|}}\sinh\left[
\frac{1}{3}\mbox{arcsinh}\left\{3\sqrt{|\delta|}(1-e^{-2\Phi_\infty}b)\right\}
\right]\quad\mbox{for~}\delta<0\,.
\end{equation}
The function $F(b)$ defined in the previous section can be rewritten as
\begin{equation}
F(b)=F_0(1-f\varphi(b))\,,
\label{5_7}
\end{equation}
and then we find that the function $\varphi(b)$ satisfies $\varphi'(0)=1$.
Substituting the `solution' into the equation of motion, we get
\begin{equation}
f\varphi(b)=Y(b)+\frac{2f b}{3\left[1+h(2 Y(b)-Y(b)^2)\right]}\,,
\label{5_8}
\end{equation}
where
\begin{eqnarray}
Y(b)&=&1-\frac{2\sqrt{1+h}}{\sqrt{h}}\sin\left[
\frac{1}{3}\arcsin\left\{\frac{3\sqrt{h}
\left(1+\frac{2}{3}h\right)}{2(1+h)^{3/2}}
\left(1-\frac{f}{3+2h}b\right)
\right\}\right]\quad\mbox{for~}h>0\,,\label{5_9}\\
Y(b)&=&1-\frac{2\sqrt{1+h}}{\sqrt{|h|}}\sinh\left[
\frac{1}{3}\mbox{arcsinh}\left\{\frac{3\sqrt{|h|}
\left(1+\frac{2}{3}h\right)}{2(1+h)^{3/2}}
\left(1-\frac{f}{3+2h}b\right)
\right\}\right]\quad\mbox{for~}h<0\,.\label{5_10}
\end{eqnarray}

\begin{figure}[ht]
\centering
\includegraphics[height=3.5cm]
{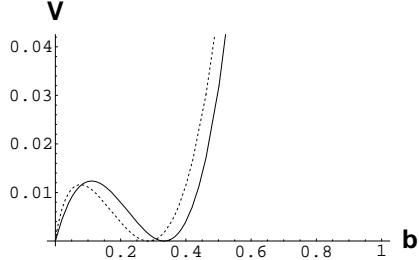}
\caption{%
The shape of the scalar potential is plotted against $b=e^{2\Phi}$ (only
shown for small $b$). The rigid line exhibits the case with $h=0$. The
dotted line  exhibits the case with $h=\infty$.}
\label{fig2}
\end{figure}

In the limit of $h\rightarrow 0$, we obtain $f\varphi(b)=fb$ as we
have already  known. On the other hand, another limiting case
$h\rightarrow\infty$ yields, if we set $f=3+2h$ for further simplicity,
$r_g=\frac{4GM}{3}$ and one can find 
\begin{equation}
b=1-\frac{2GM}{r}+\frac{4}{27}\frac{(2GM)^3}{r^3}
=1-\frac{3r_g}{2r}+\frac{r_g^3}{2r^3}
=\frac{3}{2}y^2-\frac{1}{3}y^3\,.
\end{equation}
The shape of the potential is shown in FIG.~\ref{fig2}.
Unfortunately the potential varies very slightly with $h$ in the present
assumption for parametrization.

\subsection{An inverse problem on a spherical body---
pseudo-gravastar}
\label{5_2}
Next, we consider a spherical configuration of a gravitating perfect
fluid. We assume that the outside of the radius of such a `star' is
empty vacuum. 
The vacuum solutions and the corresponding scalar potential have been
assumed and found in the previous subsection.

We assume that the energy-momentum tensor is given
in the form of a perfect fluid
\begin{equation}
T_{\mu\nu}=(\rho+p)u_\mu u_\nu-p q_{\mu\nu}\,,
\end{equation}
where $u_\mu$ is the four-velocity which satisfies $q_{\mu\nu}u^\mu
u^\nu=1$,
$\rho$ is the energy density and
$p$ is the pressure of the fluid. In the static case, the four-velocity
has only the time-like component.
Then, the quantities in (\ref{ETC}) read
\begin{equation}
T=\rho-3p\,,\quad E=-p\,,\quad C^\lambda=0\,.
\end{equation}
The dynamical equation (\ref{DE}) is expressed, with the metric
(\ref{stm}), as 
\begin{eqnarray}
& &\sqrt{\frac{V(\Phi(r))}{A(r)B(r)}}\frac{1}{r^2}\frac{d}{dr}
\left[\sqrt{\frac{B(r)}{A(r)}}r^2\frac{d\Phi(r)}{dr}\right]\nonumber \\
&=&\frac{V(\Phi)}{r^2e^{3\Phi}|1-r\Phi'|}\left[
\frac{\sqrt{V(\Phi)}r^2\Phi'}{e^\Phi|1-r\Phi'|}\right]'=
\frac{\kappa}{2}\left[\rho(r)-
\left(5-\frac{1}{2V}\frac{dV}{d\Phi}\right)p(r)\right]
\,.
\label{eq1}
\end{eqnarray}
where the prime
${}'$ denotes the derivative with respect to the radial coordinate $r$.
On the other hand, the conservation of the energy-momentum tensor 
leads to the equation
\begin{equation}
\nabla_\mu T^{\mu\nu}=\frac{1}{\sqrt{-q}}\partial_\mu
(\sqrt{-q}\,T^{\mu\nu})+\Gamma^\nu_{\rho\sigma}T^{\rho\sigma}=0\,,
\end{equation}
where
\begin{equation}
\Gamma^\nu_{\rho\sigma}=\frac{1}{2}q^{\nu\lambda}(\partial_\rho 
q_{\lambda\sigma}+\partial_\sigma 
q_{\lambda\rho}-\partial_\lambda 
q_{\rho\sigma})\,.
\end{equation}
For the perfect fluid, the equation of conservation becomes
\begin{equation}
p'(r)+[\rho(r)+p(r)]\Phi'(r)=0\,.
\label{eq2}
\end{equation}
The equations (\ref{eq1}) and (\ref{eq2}), together with a certain
equation of state, which gives the relation between $\rho$ and $p$,
completely governs the structure of the spherical body in GSG.
This is a direct problem on the structure of a `star'.

Solving the equations is straightforward but tedious in genaral cases. We
leave the study on the generic solutions for various equation of state for
future work. Here, we assume, the smooth metric inside the `star' and we
investigate  the behaviors of the energy density and the pressure as
another inverse problem.

In this study, we simply assume that the $(00)$ component of the metric
inside a spherical star can be expressed by
\begin{equation}
q_{00}=C_0-C_2 r^2\,,
\label{dS}
\end{equation}
where $C_0$ and $C_2$ are constants.
Incidentally, this behavior matches the de Sitter space in GR.
Mazur and Mottla proposed `gravastars', which have the de Sitter
spacetime structure inside them, about a decade ago \cite{MM}.
Therefore, the present analysis can be considered as the first step to
investigate such simple and interesting exotic objects, say,
`pseudo-gravastars' in GSG (and similar theories mentioned in
Sec.~\ref{sec1}).


Although the gravastars in GR may have a shell of an exotic matter at
their surface, we take a simple assumption in this paper, for the
`pseudo-gravastars' in GSG;
We define the surface of a star as
$p(r_*)=0$, where $r_*$ is the radius of the star. 

We use the result of
the previous subsection, so we adopt $V(\Phi)=e^{2\Phi}F(e^{2\Phi})^2$,
where $F$ is defined by (\ref{5_7}--\ref{5_10}) with the choice $f=3+2h$. 
The metric for the exterior of the star can be described by 
\begin{equation}
b(r)=e^{2\Phi(r)}=b_e(r)=1-\frac{2GM}{r}+\frac{4\delta}{3}\frac{(2GM)^3}{r^3}\,,\quad
\mbox{for }r>r_*\,.
\end{equation}
On the other hand, we assume that the metric is given by the function
(\ref{dS}) which smoothly connects to $b_e(r)$:
\begin{equation}
b(r)=b_*(r)\equiv
b_e(r_*)-\frac{b_e'(r_*)}{2}\left(1-\frac{r^2}{r_*^2}\right)\equiv
e^{2\Phi_*(r)}\,,\quad
\mbox{for }r<r_*\,.
\end{equation}
For $r<r_*$, eliminating $\rho$ from Eqs.~(\ref{eq1},\ref{eq2}) yields
\begin{equation}
\left[e^{6\Phi_*(r)}\frac{1}{\sqrt{V(\Phi_*(r))}}\kappa
p(r)\right]'
=-\frac{2
e^{3\Phi_*(r)}\sqrt{V(\Phi_*(r))}\Phi_*'(r)}{r^2(1-r\Phi_*'(r))}\left[
\frac{\sqrt{V(\Phi_*(r))}r^2\Phi_*'(r)}{e^{\Phi_*(r)}(1-r\Phi_*'(r))}\right]'
\,,
\end{equation}
for $r<r_*$. To obtain the solution for $p(r)$, we have only to integrate
the equation.

\begin{figure}[ht]
\centering
\includegraphics[height=3.5cm]
{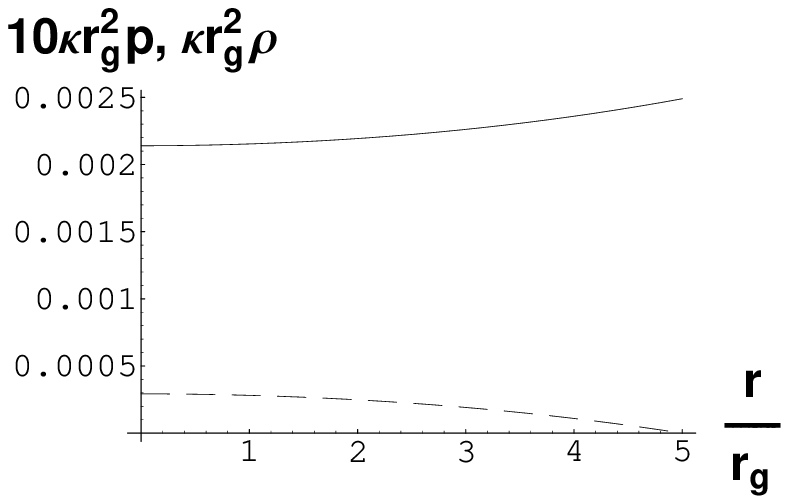}
\caption{%
The energy density (solid line) and the pressure (dashed line) are
plotted against $r/r_g$ in the case with the parameter $h=0$ and $r_*=5
r_g$.}
\label{fig3}
\end{figure}
\begin{figure}[ht]
\centering
\includegraphics[height=3.5cm]
{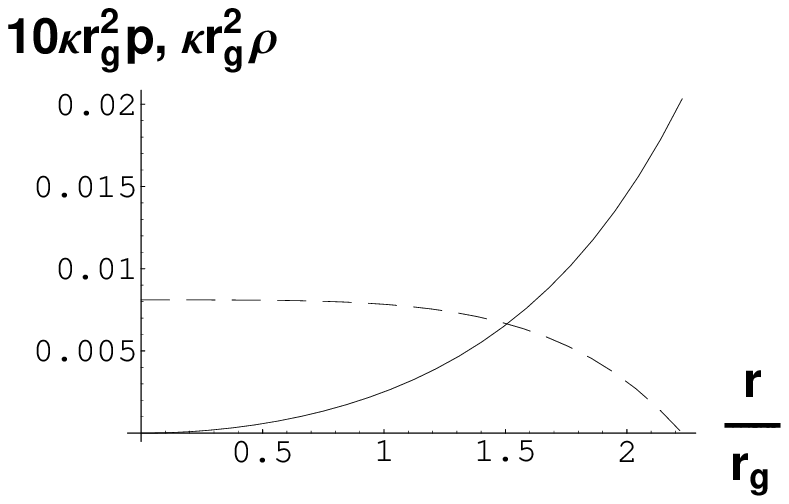}
\caption{%
The energy density (solid line) and the pressure (dashed line) are
plotted against $r/r_g$ in the case with the parameter $h=0$ and
$r_*=2.228 r_g$.}
\label{fig4}
\end{figure}

\begin{figure}[ht]
\centering
\includegraphics[height=3.5cm]
{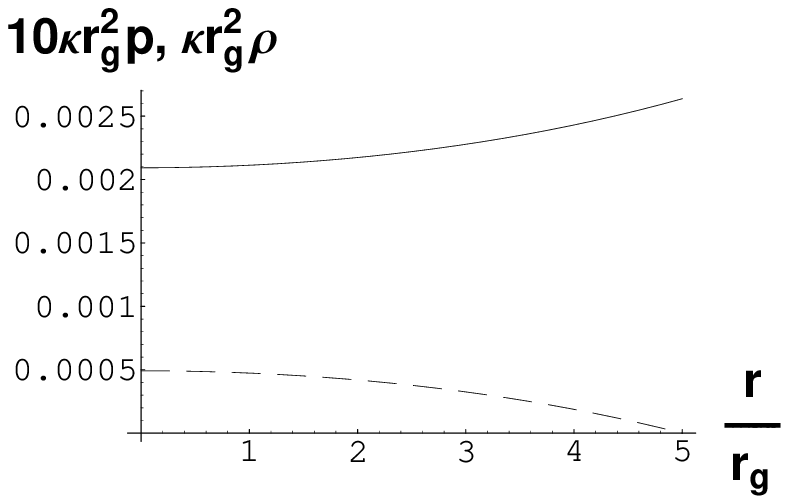}
\caption{%
The energy density (solid line) and the pressure (dashed line) are
plotted against $r/r_g$ in the case with the parameter $h=\infty$ and
$r_*=5 r_g$.}
\label{fig5}
\end{figure}
\begin{figure}[ht]
\centering
\includegraphics[height=3.5cm]
{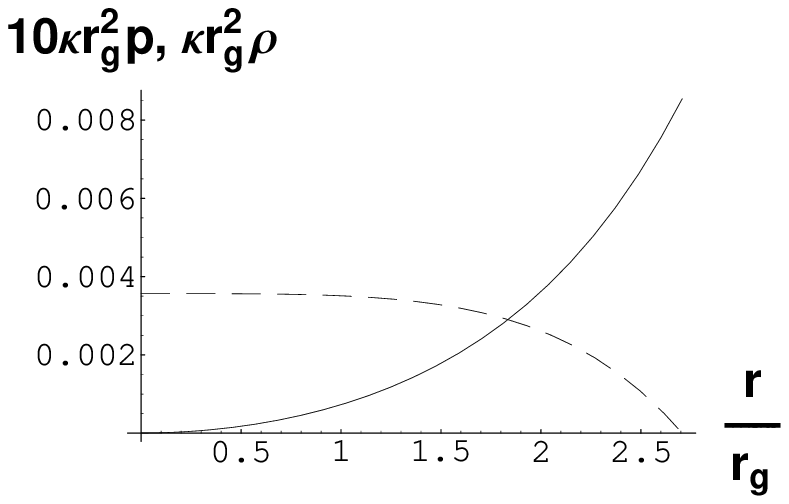}
\caption{%
The energy density (solid line) and the pressure (dashed line) are
plotted against $r/r_g$ in the case with the parameter $h=\infty$ and
$r_*=2.707 r_g$.}
\label{fig6}
\end{figure}

Some numerical examples are shown in
FIGs.~\ref{fig3}, \ref{fig4}, \ref{fig5} and \ref{fig6}.
In each figure, $10\kappa r_g^2 p$ and  $\kappa r_g^2 \rho$ are plotted
against $r/r_g$. FIGs.~\ref{fig3} and \ref{fig4} are in the case with the
parameter $h=0$ while FIGs.~\ref{fig5} and
\ref{fig6} are in the case with $h=\infty$.
In FIGs.~\ref{fig3} and \ref{fig5}, we choose $r_*=5 r_g$.
FIGs.~\ref{fig4} and \ref{fig6} indicate the results with $\rho(0)=0$.

In both cases of $h=0$ and $h=\infty$, the behaviors of $\rho$ and $p$
are  qualitatively equal respectively. 
In FIGs.~\ref{fig4} and \ref{fig6}, the values for $r_*$ are considered as
the lower bound for the radii of the stars of the present type.
Below the critical value for the radius, the energy density at the
origin becomes negative.

In contrast to the gravastars in GR, we find that the value of the 
pressure $p$ can be positive and simply decreasing in the present
pseudo-gravastar in GSG. Of course, the geometry of the interior of the
star is not an exact de Sitter space, for $q_{11}\ne -(q_{00})^{-1}$ in
GSG.

The `stars' constructed above are, however, not realistic
because the energy density $\rho$ also should decrease along with $r$
for a normal equation of state.%
\footnote{As for a possible improvement, we should take shell
boundaries and anisotropic pressures for exotic stars into account.} 
Nevertheless, we suppose that the characteristic scales of the `star' in
GSG revealed in the simple approach are useful for setting the starting
values of numerical calculation in future work.

\section{Summary and prospects}
\label{sec7}

In this paper, we considered the geometric scalar gravity of Novello and
collaborators and found conditions of the scalar potential that had a
permissible solution in the case of weak and strong gravity. We found
that the simple potential of Novello et al.~leads to the solution
which agrees with the Schwarzschild solution of GR
and small corrections to the potential are permitted and the
generic potential leads to the `black hole solution' with a modified
mass-horizon radius relation (i.e., $r_g\ne 2GM$), of which asymptotic
behavior describes weak gravity up to the post-Newtonian order.
We also showed possible solutions for `pseudo-gravastars' in GSG with
generic scalar potentials.

We will try to extend the geometric scalar theory of gravity
with additional degrees of freedom.
We also wish to study the mass-radius relation (known in GR
\cite{Buchdahl}) which may exist in a solution describing a spherical
star in GSG by numerical analyses. In addition, we want to study
`cosmologies' expressed by the generic scalar potentials in geometric
scalar gravity. We hope to report soon on these issues.

The quantization of GSG is another very
interesting subject. We wish to consider it in the future research.

\bigskip

\noindent
{\bf Note added:}

After completion of this manuscript, we have noticed a new preprint
arXiv:1508.02665 [gr-qc] by Jardim and Landim \cite{JL} on GSG.
It discusses the issue of the cosmological constant in GSG and argues
an `inverse problem' to realize the de Sitter spacetime in GSG.

\acknowledgments
We thank Prof. Kumar Shwetketu Virbhadra for valuable information on
naked singularities.

We also thank Prof. Sunny Vagnozzi for providing information on their
work on modified gravity including scalar derivative terms
\cite{MSV,MSVZ}.



\bibliographystyle{apsrev4-1}




\end{document}